\documentclass{article}
\usepackage[utf8]{inputenc}
\usepackage[T1]{fontenc}
\usepackage{graphicx}
\usepackage{amsmath,amssymb,amsfonts}%
\usepackage{a4wide}
\usepackage[round]{natbib}
\usepackage{authblk}
\usepackage{hyperref}

\title{The generalised balanced power diagram: flat sections, affine transformations and an improved rendering algorithm}
\author{Felix Ballani\footnote{Email address: \texttt{f.ballani@hzdr.de} | ORCID-iD: \href{https://orcid.org/0000-0002-0415-0220}{0000-0002-0415-0220}}}
\affil{Helmholtz-Zentrum Dresden -- Rossendorf, Helmholtz Institute Freiberg for Resource Technology, Chemnitzer Str. 40, 09599 Freiberg, Germany}
\date{}
\begin{document}

\maketitle

\begin{abstract}
The generalised balanced power diagram (GBPD) is regarded in the literature as a suitable geometric model for describing polycrystalline microstructures with curved grain boundaries. This article compiles properties of GBPDs with regard to affine transformations and flat sections. Furthermore, it extends an algorithm known for power diagrams for generating digital images, which is more efficient than the usual brute force approach, on GBPDs.
\end{abstract}
%
\section{Introduction}
This article deals with a type of tessellation of $\mathbb{R}^d$ (or some subset) that has been introduced and applied in the literature over the past decade at least, both under the name generalised balanced power diagram (GBPD), see \citet{AlpersEtAl2015,TeferraGraham2015,SedivyEtAl2016,SedivyEtAl2017,PetrichEtAl2021}, and anisotropic power diagram (APD), see \citet{BuzeEtAl2024}. This is a \emph{Voronoi}-like tessellation generated from a finite set of seed points in $\mathbb{R}^d$, i.\,e. each point in space is assigned to the seed point closest to it in terms of a specific given distance measure. For a GBPD, the distance measure is specified as follows:
\begin{equation}
\mathrm{dist}\big(x,s\big)=\mathrm{dist}_G(x,s,M,w):=(x-s)^{\top}M(x-s)-w,\quad x\in\mathbb{R}^d,
\end{equation}
where $s\in\mathbb{R}^d$ is a seed point, $M$ is a positive definite matrix of $d\times d$ real values and $w\in\mathbb{R}$.

The majority of literature to date has focused on the question of how GBPDs can be reconstructed, i.\,e. how a polycrystalline microstructure given as a voxel image can best be approximated as a concrete GBPD \citep{AlpersEtAl2015,TeferraGraham2015,SedivyEtAl2016,SedivyEtAl2017,SedivyEtAl2018, TeferraRowenhorst2018,PetrichEtAl2021,AlpersEtAl2023,BuzeEtAl2024,AlpersEtAl2025}. Furthermore, \citet{JungRedenbach2024} give an analytic representation of the vertices and edges of the cells of the GBPD in 2D.

This article focuses on the question of how digital images of GBPDs can be generated as efficiently as possible from an algorithmic point of view -- performance is typically achieved through parallelisation and computing on GPUs, see e.\,g. \citet{BuzeEtAl2024} -- and how GBPDs behave under flat sections. In the light of the identifiability of the determining pieces of a GBPD, the behaviour with regard to affine transformations is also described. All these aspects can play a role when modelling polycrystalline microstructures using stochastic models based on the GBPD concept; however, such stochastic modelling is not part of this paper.

The article is structured as follows. After providing some necessary definitions in Section \ref{sec:def}, Section \ref{sec:trafo} deals with the behaviour of GBPDs with regard to certain affine transformations. Section \ref{sec:section} examines the behaviour of GBPDs under flat sections. In Section \ref{sec:render}, an improved algorithm for rendering digital images is transferred to GBPDs, and then in Section \ref{sec:Poisson} its complexity is discussed using the example of a marked Poisson process. Finally, Section \ref{sec:conclusion} concludes the article.

%
\section{Definition}\label{sec:def}
Let $\psi$ be a countable set of marked points of the form $(s,M,w)$ where $s\in\mathbb{R}^d$, $M$ is a positive definite matrix of $d\times d$ real values and $w\in\mathbb{R}$. For each $(s,M,w)\in\psi$ the (closed) set
\begin{equation}
C((s,M,w),\psi) := \{x\in\mathbb{R}^d: \mathrm{dist}_G(x,s,M,w)\leq\mathrm{dist}_G(x,s',M',w')\text{ for all }(s',M',w')\in\psi\}
\end{equation}
defines a \emph{cell} and $s$ is called its \emph{seed}, $M$ is called its \emph{anisotropy matrix}, and $w$ is called its \emph{(Laguerre) weight}; the triple $(s,M,w)$ is henceforth referred to as its \emph{generator}. The collection of all non-empty cells of $\psi$,
\begin{equation}
\mathcal{G}(\psi):=\{C((s,M,w),\psi): (s,M,w)\in\psi,\,C((s,M,w),\psi)\neq\varnothing\}  
\end{equation}
is a \emph{tessellation} of $\mathbb{R}^d$ \citep{Moller1989}, i.\,e., the interiors of the cells are disjoint, the cells are space-filling and every bounded subset of $\mathbb{R}^d$ is hit by only finitely many cells. This tessellation $\mathcal{G}(\psi)$ is called the \emph{generalised balanced power diagram (GBPD)} or anisotropic power diagram generated by $\psi$. $\psi$ is therefore occasionally referred to as the \emph{generator} of $\mathcal{G}(\psi)$ in the following.

The GBPD extends several tessellation classes. In case that all anisotropy matrices $M$ from $\psi$ are equal to the identity matrix and all weights $w$ are equal to the same value, the GBPD coincides with a Voronoi tessellation/diagram \citep[see, e.\,g.,][]{OkabeEtAl2000} of the generator set $\{s:(s,M,w)\in\psi\}$. In case that all anisotropy matrices $M$ from $\psi$ are equal to the identity matrix, the GBPD coincides with a Laguerre tessellation/power diagram \citep{Aurenhammer1987,Lautensack2007,LautensackZuyev2008} of the generator set $\{(s,w):(s,M,w)\in\psi\}$. Finally, in case that all weights $w$ are equal to the same value, the GBPD coincides with an anisotropic Voronoi diagram \citep{LabelleShewchuk2003,Jeulin2013,AltendorfEtAl2014,NulandEtAl2021} of the generator set $\{(s,M):(s,M,w)\in\psi\}$.

\begin{figure}[t]
\centering
\includegraphics[width=0.45\textwidth]{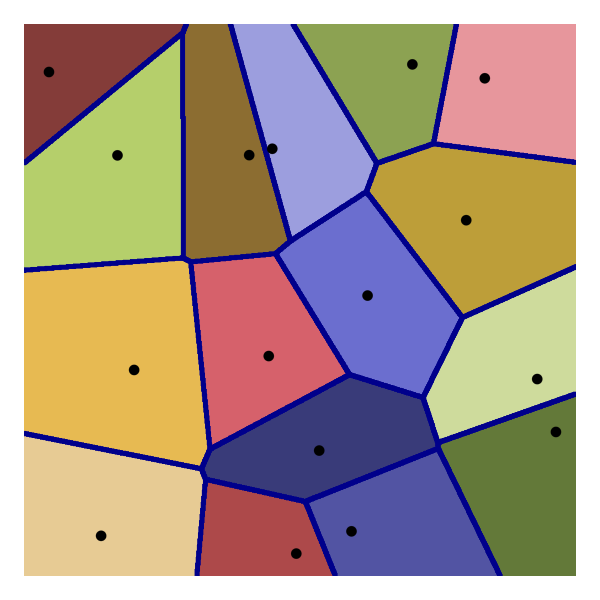}
\includegraphics[width=0.45\textwidth]{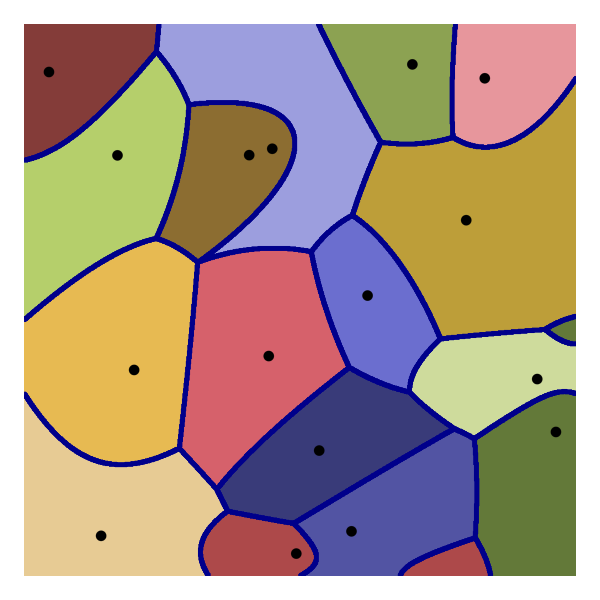}
\caption{Laguerre tessellation (left) and GBPD (right) with respect to the same set of seed points (black dots) and each the same weights. Even in Laguerre tessellations there might be seeds with an empty cell or lying outside their cell. In GBPDs cells can be non-convex and even disconnected.}
\label{fig:gbpd}    
\end{figure}

Figure \ref{fig:gbpd} shows the typical appearance of a GBPD; it generally has curved cell boundaries, whereas a Laguerre tessellation always has flat cell boundaries.

The generator $\psi$ of a GBPD $\mathcal{G}(\psi)$ is not unique. For all $v\in\mathbb{R}$ the generators $\{(s,M,w+v):(s,M,w)\in\psi\}$ result in the same GBPD since all distances 
\begin{equation}\label{eqn:weight:shift}
\mathrm{dist}_G(x,s,M,w+v) = \mathrm{dist}_G(x,s,M,w) + v 
\end{equation}
are shifted by the same amount. In a similar fashion for all $a>0$ the generators
$\{(s,aM,aw):(s,M,w)\in\psi\}$ result in the same GBPD since all distances 
\begin{equation}\label{eqn:weight:scale}
\mathrm{dist}_G(x,s,aM,aw) = a\cdot\mathrm{dist}_G(x,s,M,w) 
\end{equation}
are scaled by the same amount.

%
\section{Transformations}\label{sec:trafo}
Under various transformations, the property of a tessellation being a GBPD does not change. This, together with (\ref{eqn:weight:shift}) and (\ref{eqn:weight:scale}), may be of interest when it comes to the identifiability of model parameters in stochastic models with GBPDs. In the following, let $\psi$ be a generator of the GBPD $\mathcal{G}(\psi)$.
%
\subsection{Translation}\label{subsec:translation}
For any $y\in\mathbb{R}^d$ the translated GBPD $\mathcal{G}(\psi)+y$ is a GBPD generated by
\[
\psi+y:=\{(s+y,M,w):(s,M,w)\in\psi\}.
\]
%
\subsection{Rotation}\label{subsec::rotation}
Let $U$ be any rotation matrix in $\mathbb{R}^d$. Due to
\begin{align*}
&UC((s,M,w),\psi)\\ 
&= \{Ux\in\mathbb{R}^d: (x-s)^{\top}M(x-s)-w\leq(x-s')^{\top}M'(x-s')-w'\;\forall\;(s',M',w')\in\psi\}\\
&= \{y\in\mathbb{R}^d: (U^{\top}y-s)^{\top}M(U^{\top}y-s)-w\leq(U^{\top}y-s')^{\top}M'(U^{\top}y-s')-w'\;\forall\;(s',M',w')\in\psi\}\\
&= \{y\in\mathbb{R}^d: (y-Us)^{\top}UMU^{\top}(y-Us)-w\leq(y-Us')^{\top}UM'U^{\top}(y-Us')-w'\;\forall\;(s',M',w')\in\psi\}
\end{align*}
the rotated GBPD $U\mathcal{G}(\psi)$ is a GBPD generated by
\[
U\psi:=\{(Us,UMU^{\top},w):(s,M,w)\in\psi\}.
\]
%
\subsection{Scaling}\label{subsec:scaling}
Let $a>0$. Due to
\begin{align*}
&aC((s,M,w),\psi)\\
&=\{y\in\mathbb{R}^d: (y-as)^{\top}(a^{-2}M)(y-as)-w\leq(y-as')^{\top}(a^{-2}M')(y-as')-w'\;\forall\;(s',M',w')\in\psi\}\\
&=\{y\in\mathbb{R}^d: (y-as)^{\top}M(y-as)-a^2w\leq(y-as')^{\top}M'(y-as')-a^2w'\;\forall\;(s',M',w')\in\psi\}
\end{align*}
the scaled GBPD $a\mathcal{G}(\psi)$ is a GBPD generated by
\[
\{(as,a^{-2}M,w):(s,M,w)\in\psi\}\text{ and }\{(as,M,a^2w):(s,M,w)\in\psi\}.
\]
%
\subsection{Linear distortion}\label{subsec:distortion}
Let $A$ be an invertible $d\times d$ matrix of real values. Then (more generally, and including cases \ref{subsec::rotation} and \ref{subsec:scaling}) the transformed GBPD $A\mathcal{G}(\psi)$ is a GBPD generated by
\[
A\psi:=\{(As,A^{-\top}MA^{-1},w):(s,M,w)\in\psi\}.
\]

%
\section{Flat sections of GBPDs}\label{sec:section}
Although scanning methods that provide three-dimensional images are often used today in the examination of natural or technical microstructures, methods based on flat sections are still in use, for example for reasons of speed. The following therefore outlines how GBPDs behave under flat sections. As before, let $\psi$ be a generator of the ($d$-dimensional) GBPD $\mathcal{G}(\psi)$.
%
\subsection{Section with a hyperplane}\label{subsec:hyperplane}
Let $H$ be a hyperplane of $\mathbb{R}^d$. Due to \ref{subsec::rotation} we may assume that $H$ is of the form $$H=\{x=(x_1,\ldots,x_d)^{\top}\in\mathbb{R}^d:\,x_k=h\}$$ for some $h\in\mathbb{R}$ and some $k\in\{1,\ldots,d\}$. The object of interest is the tessellation 
\begin{equation}
\mathcal{G}(\psi)\cap H:=\{C\cap H:\,C\in\mathcal{G}(\psi)\}. 
\end{equation}
If $H$ is not of the assumed particular form, then there is some rotation matrix $U$ such that $UH$ is of that form and the resulting sectional tessellation is $U^{\top}(\mathcal{G}(U\psi)\cap UH)$.

In what follows, for any (column) vector $x=(x_1,\ldots,x_d)^{\top}\in\mathbb{R}^d$ denote by $x_{\not k}$ the (column) vector consisting of all components of $x$ except the $k$-th component $x_k$. Similarly, for some matrix $M\in\mathbb{R}^{d\times d}$ let $M_{\not k,k}=(M_{i,k})_{i\neq k}$ and $M_{\not k,\not k}=(M_{i,j})_{i\neq k,j\neq k}$.

Let $(s,M,w)\in\psi$. For each $x\in\mathbb{R}^d$ the distance $\mathrm{dist}_G(x,s,M,w)$ can be decomposed in the following way, using $y_{\not k}=(x_k-s_k)M_{\not k,\not k}^{-1}M_{\not k,k}$, the symmetry of $M$ and the property that because $M$ is positive definite, $M_{\not k,\not k}$ is also positive definite and therefore invertible:
\begin{align*}
\mathrm{dist}_G(x,s,M,w)&=(x-s)^{\top}M(x-s)-w\\
&=(x_{\not k}-s_{\not k})^{\top}M_{\not k,\not k}(x_{\not k}-s_{\not k})+2(x_k-s_k)M_{\not k,k}^{\top}(x_{\not k}-s_{\not k})+(x_k-s_k)^2M_{k,k}-w\\
&=(x_{\not k}-s_{\not k}+y_{\not k})^{\top}M_{\not k,\not k}(x_{\not k}-s_{\not k}+y_{\not k})+(x_k-s_k)^2(M_{k,k}-M_{\not k,k}^{\top}M_{\not k,\not k}^{-1}M_{\not k,k})-w
\end{align*}
Consequently, within the hyperplane $H=\{x:\,x_k=h\}$, the distance is again a distance of type $\mathrm{dist}_G$,
\begin{equation}
\mathrm{dist}_G\big(x_{\not k},\tilde{s},\tilde{M},\tilde{w}\big)=(x_{\not k}-\tilde{s})^{\top}\tilde{M}(x_{\not k}-\tilde{s})-\tilde{w},
\end{equation}
with respect to the seed
\begin{equation}
\tilde{s}=s_{\not k}-(h-s_k)M_{\not k,\not k}^{-1}M_{\not k,k},
\end{equation}
the anisotropy matrix
\begin{equation}
\tilde{M}=M_{\not k,\not k}, 
\end{equation}
and the weight
\begin{equation}
\tilde{w}=w-(h-s_k)^2(M_{k,k}-M_{\not k,k}^{\top}M_{\not k,\not k}^{-1}M_{\not k,k}).
\end{equation}

%
\subsection{Section with a $q$-dimensional affine space, $1\leq q\leq d-2$}
In case $d\geq3$ consider a $q$-dimensional affine subspace, $1\leq q\leq d-2$, w.l.o.g. assumed to be of the form $H_K=\{x=(x_1,\ldots,x_d)^{\top}\in\mathbb{R}^d:\,x_k=h_k,\,k\in K\}$ for some $K\subset\{1,\ldots,d\}$, $|K|=d-q$, and some $h_k\in\mathbb{R}$, $k\in K$. Similarly to the previous section, denote by $x_{K}$ and $x_{\not K}$ the vector consisting of all components $x_k$ of $x\in\mathbb{R}^d$ for $k\in K$, and $k\notin K$, respectively, and, for some matrix $M\in\mathbb{R}^{d\times d}$, let $M_{K,K}=(M_{i,j})_{i\in K,j\in K}$, $M_{\not K,K}=(M_{i,j})_{i\notin K,j\in K}$, and $M_{\not K,\not K}=(M_{i,j})_{i\notin K,j\notin K}$.

Analogously, within $H_K$, the distance is again a distance of type $\mathrm{dist}_G$,
\begin{equation}
\mathrm{dist}_G\big(x_{\not K},\tilde{s},\tilde{M},\tilde{w}\big)=(x_{\not K}-\tilde{s})^{\top}\tilde{M}(x_{\not K}-\tilde{s})-\tilde{w},
\end{equation}
with respect to the seed
\begin{equation}
\tilde{s}=s_{\not K}-(h_K-s_K)^{\top}M_{\not K,\not K}^{-1}M_{\not K,K},
\end{equation}
the anisotropy matrix
\begin{equation}
\tilde{M}=M_{\not K,\not K}, 
\end{equation}
and the weight
\begin{equation}
\tilde{w}=w-(h_K-s_K)^{\top}(M_{K,K}-M_{\not K,K}^{\top}M_{\not K,\not K}^{-1}M_{\not K,K})(h_K-s_K).
\end{equation}

%
\section{Rendering of digital images}\label{sec:render}
A common task for tessellations is their representation in the form of a digital image. For each point $x$ from a finite discrete set of points $X$ (typically pixels or voxels), where $X\subset W$ for some convex compact simulation window $W\subset\mathbb{R}^d$, it must be decided to which cell of the tessellation $x$ belongs. In the case of Voronoi and Laguerre tessellations, all cells are convex polytopes whose respective determining pieces can be calculated from the generators of the tessellation. Then a subsequent membership evaluation of the points of $X$ is easy to perform. However, GBPDs generally have curved cell boundaries. This makes them interesting for use in modelling microstructures. However, even if analytical descriptions of cell boundaries are known \citep{JungRedenbach2024}, it is more difficult to assess cell membership based on these descriptions, as the cells generally do not have to be convex or topologically connected (see also Figure \ref{fig:gbpd}). It may therefore be necessary to determine the cell memberships directly from the generators via the corresponding distances. 

%
\subsection{The ‘brute force’ algorithm}
The simplest algorithm, sometimes referred to as the ‘brute force’ algorithm, consists of calculating the distance $\mathrm{dist}_G(x,s,M,w)$ for each $x\in X$ for all generators $(s,M,w)\in\psi$ and thus determining the nearest seed $s$. If $\psi$ contains $n$ generators and $X$ consists of $N$ points, this algorithm has a complexity of $\mathcal{O}(Nn)$ \citep{NulandEtAl2021,Moulinec2022}. 

%
\subsection{The algorithm of \citet{Moulinec2022} adapted to GBPDs}\label{subsec:algorithm}
%
\begin{figure}[t]
\centering
\includegraphics[width=0.32\textwidth]{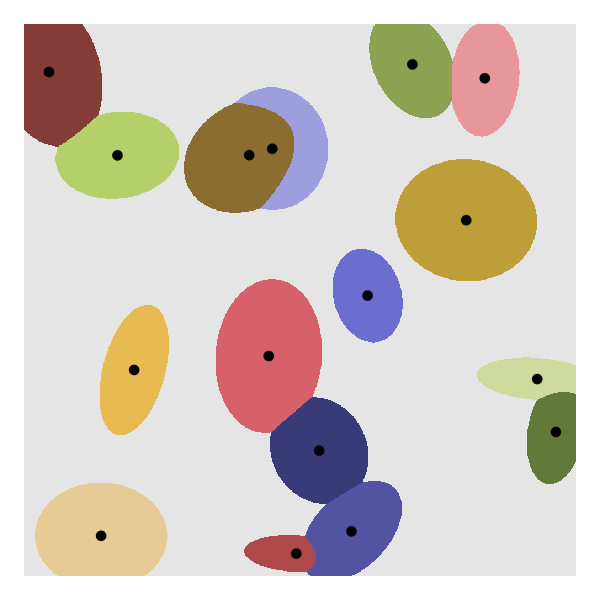}
\includegraphics[width=0.32\textwidth]{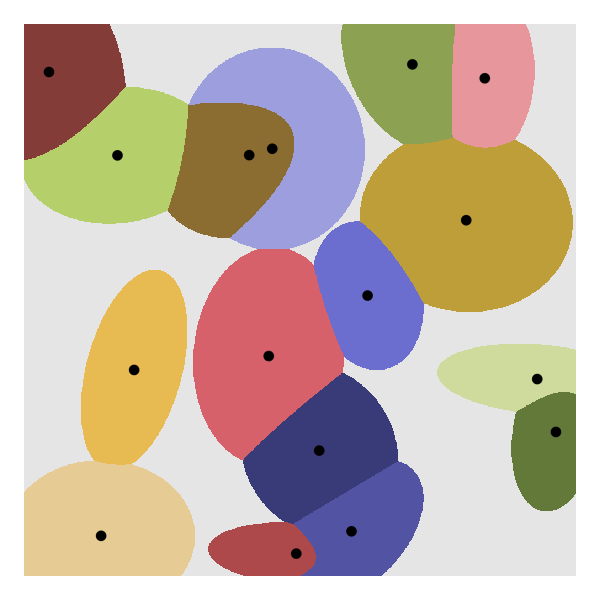}
\includegraphics[width=0.32\textwidth]{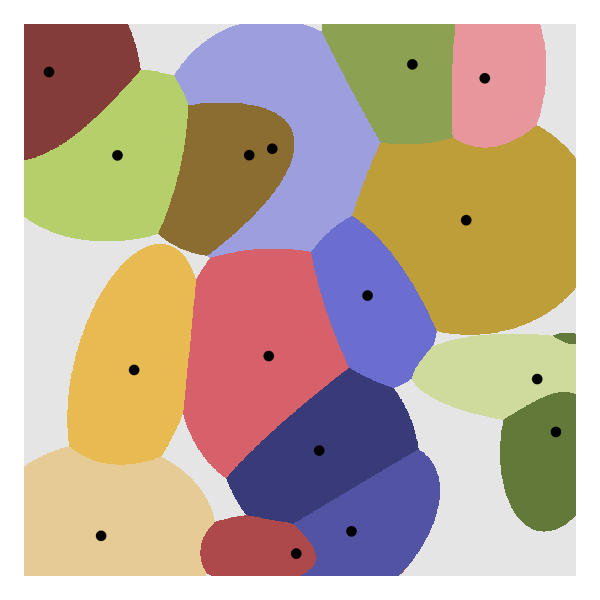}
\caption{Assigned memberships to seed points after the first step of the improved algorithm for three different distances $t$ (growing from left to right). For each pixel farther than $t$ away from any seed point, i.\,e. outside the coloured region, the 'brute force' algorithm has to be applied.}
\label{fig:step1}    
\end{figure}
\citet{Moulinec2022} proposes an alternative algorithm with a lower complexity of $\mathcal{O}(N\log(n))$ for Voronoi and Laguerre tessellations. This section explains how this alternative algorithm can be extended to GBPDs and affine sections of GBPDs. 

The basic principle of the improved algorithm is to first examine a specific neighbourhood of all seed points and determine for each $x\in X$ in this neighbourhood whether the seed under consideration is the closest, and then, in a second step, process the remaining points $x\in X$ (i.\,e. that have not yet been visited) using the ‘brute force’ algorithm described above.

In the following, it is assumed that all weights of the generator $\psi$ are non-negative. To determine the neighbourhood required in the first step of the improved algorithm, a global $t>0$ is specified for all seed points. The choice of $t$ has an influence on the complexity of the algorithm and is discussed in Section \ref{sec:Poisson} for a particular model.  

For each generator $(s,M,w)\in\psi$, let 
\begin{equation}
E_t(s,M,w)=\{x\in\mathbb{R}^d:\mathrm{dist}_G(x,s,M,w)\leq t\}
\end{equation}
be the set of all points in $\mathbb{R}^d$ that are at most $t$ away from the seed point $s$ in the sense of $\mathrm{dist}_G$. The neighbourhood of $s$ within $X$ to be run through for a generator $(s,M,w)$ is then precisely $X\cap E_t(s,M,w)$. Due to
\begin{equation}\label{eqn:ellipsoid}
\mathrm{dist}_G(x,s,M,w)\leq t\quad\Leftrightarrow\quad(x-s)^{\top}M(x-s)\leq t+w\quad\Leftrightarrow\quad(x-s)^{\top}\frac{M}{t+w}(x-s)\leq1
\end{equation}
$E_t(s,M,w)$ is then a $d$-dimensional ellipsoid centered in $s$, whose principal axes are given by the eigenvectors of $M/(t+w)$ and whose semi-axes are given by the reciprocals of the (positive) roots of the eigenvalues of $M/(t+w)$.

Since the membership of a point $x$ in the neighbourhood $X\cap E_t(s,M,w)$ is defined precisely by the property $\mathrm{dist}_G(x,s,M,w)\leq t$, but unnecessary calculations of the distance $\mathrm{dist}_G(x,s,M,w)$ should be avoided in the improved algorithm, the following procedure can be used, particularly in the case where $X$ is a regular grid of pixels or voxels. It is based on the property that the smallest axis-aligned bounding box of $E_t(s,M,w)$ is given by 
\begin{equation}\label{eqn:box}
B_t(s,M,w)=s+\sqrt{t+w}\,\big\{(y_1,\ldots,y_d)\in\mathbb{R}^d:|y_i|\leq\sqrt{(M^{-1})_{i,i}},\,i=1,\ldots,d\big\},    
\end{equation}
since the width of a (non-degenerate) ellipsoid $\{x\in\mathbb{R}^d:\,x^{\top}Ax\leq1\}$ in the direction $u\in\mathbb{R}^d$, $\|u\|=1$, is given by $2\sqrt{u^{\top}A^{-1}u}$ \citep[][p. 461]{BoydVandenberghe2004}.

For each generator $(s,M,w)\in\psi$ the set $X\cap B_t(s,M,w)$, which can essentially be determined on the basis of a single evaluation of $M^{-1}$, has to be run through. For each $x\in X\cap B_t(s,M,w)$ the distance $\mathrm{dist}_G(x,s,M,w)$ has to be calculated, and in case $\mathrm{dist}_G(x,s,M,w)\leq t$ the membership of $x$ may have to be updated. Figure \ref{fig:step1} illustrates the situation after the first step for several values of $t$ for a two-dimensional example.

%
\subsection{Details for the section with a hyperplane}
If the GBPD is only to be rendered within a hyperplane $H=\{x=(x_1,\ldots,x_d)^{\top}\in\mathbb{R}^d:\,x_k=h\}$, $h\in\mathbb{R}$, $k\in\{1,\ldots,d\}$ (see also Section \ref{subsec:hyperplane}), the first step of the improved algorithm can be performed as follows.

Since $B_t(s,M,w)$ is the smallest axis-aligned bounding box of the ellipsoid $E_t(s,M,w)$, the latter hits the section hyperplane $H$ with an intersection of measure larger than 0 if
\begin{equation}\label{eqn:hit}
(h-s_k)^2<(t+w)(M^{-1})_{k,k}.
\end{equation}
Using the notation of Section \ref{subsec:hyperplane} this is equivalent to
\begin{equation}
\frac{(h-s_k)^2}{(M^{-1})_{k,k}}=(h-s_k)^2(M_{k,k}-M_{\not k,k}^{\top}M_{\not k,\not k}^{-1}M_{\not k,k})< t+w,
\end{equation}
since by the Schur complement $M_{k,k}-M_{\not k,k}^{\top}M_{\not k,\not k}^{-1}M_{\not k,k}$ of $M$ \citep[see, e.\,g.][A.5.5]{BoydVandenberghe2004} we have
\begin{equation}\label{eqn:Schur}
(M^{-1})_{k,k}=(M_{k,k}-M_{\not k,k}^{\top}M_{\not k,\not k}^{-1}M_{\not k,k})^{-1}.
\end{equation}
For all $x\in H$ we have equivalently
$\mathrm{dist}_G(x,s,M,w)\leq t$, $\mathrm{dist}_G(x_{\not k},\tilde{s},\tilde{M},\tilde{w})\leq t$ and
\begin{equation}\label{eqn:ellipse}
(x_{\not k}-\tilde{s})^{\top}\frac{\tilde{M}}{t+\tilde{w}}(x_{\not k}-\tilde{s})=(x_{\not k}-\tilde{s})^{\top}\frac{M_{\not k,\not k}}{t+w-(h-s_k)^2(M_{k,k}-M_{\not k,k}^{\top}M_{\not k,\not k}^{-1}M_{\not k,k})}(x_{\not k}-\tilde{s})\leq1
\end{equation}
according to (\ref{eqn:ellipsoid}). The above condition (\ref{eqn:hit}) together with (\ref{eqn:Schur}) ensures that the denominator in (\ref{eqn:ellipse}) is positive. (\ref{eqn:ellipse}) defines just the ($(d-1)$-dimensional) ellipsoid $E_t(s,M,w)\cap H$ within $H$, i.\,e. the section ellipsoid and the generating ellipsoid within $H$ coincide. As in (\ref{eqn:box}) its extreme values with respect to the coordinate axes might be used to restrict the number of distance calculations for the generator $(s,M,w)$.

%
\section{Complexity estimation for a marked Poisson process}\label{sec:Poisson}
In what follows, we assume that the generator $\psi$ of the GBPD $\mathcal{G}(\psi)$ is a realization of an independently marked stationary Poisson point process $\Psi$. That is, the set of seeds $s$ builds a homogeneous Poisson point process in $\mathbb{R}^d$ \citep{ChiuEtAl2013} of some intensity $\lambda$, $0<\lambda<\infty$, and independently each seed is marked with a random mark $(M,w)$ consisting of an anisotropy matrix $M$ and a weight $w\geq0$.

For each $t>0$ each mark $(M,w)$ is related to an associated ellipsoid $E_t(M,w):=E_t(o,M,w)$ centred in the origin $o$. The shifted ellipsoid $E_t(M,w)+s$ represents just the domain $E_t(s,M,w)$ within which the membership of points must be evaluated in the first step of the above rendering algorithm (see Section \ref{subsec:algorithm}). Provided that the condition \citep[(4.14)]{SchneiderWeil2008}
\begin{equation}\label{eqn:condition}
\mathsf{E}[V_d(E_t(M,w)\oplus B^d)] < \infty
\end{equation}
is satisfied, where $\mathsf{E}[\cdot]$ denotes expectation, $V_d(\cdot)$ the $d$-volume, $\oplus$ Minkowski addition and $B^d$ the $d$-dimensional unit ball, 
\begin{equation}
\Psi_t = \{(s,E_t(M,w)):\,(s,M,w)\in\Psi\}
\end{equation}
is a stationary Poisson germ-grain process with typical grain $E_t(M,w)$. To ensure the above condition (\ref{eqn:condition}), we henceforth assume that all weights $w$ are bounded above by $w_{\max}<\infty$ and that there exists a $r_{\max}$, $0<r_{\max}<\infty$, such that $\{x\in\mathbb{R}^d:\,x^{\top} Mx\leq1\}\subseteq r_{\max}B^d$ holds for all anisotropy matrices $M$; together, this means that $E_t(M,w)\subseteq \sqrt{t+w_{\max}}\,r_{\max}B^d$ holds.

%
\subsection{Simplified complexity estimation}\label{subsec:simplified}
First, we consider the case where, during the first step of the rendering algorithm in Section \ref{subsec:algorithm}, distance calculations are only performed within the ellipsoids $E_t(s,M,w)=s+E_t(M,w)$. This corresponds to the approach in \citet{Moulinec2022}, but is not entirely correct, since the distance for a point $x$ has to be determined even if $x$ belongs to the axis-aligned bounding box $B_t(s,M,w)$ of $E_t(s,M,w)$, but not to $E_t(s,M,w)$, at least if the decision $x\in E_t(s,M,w)$ cannot be made in any other way.

For each point $x\in X$ the number of distance calculations during the first step of the algorithm is equal to 
\begin{equation}\label{eqn:n1}
n_1(x)=\sum_{(s,E_t)\in\Psi_t}\mathbf{1}_{E_t+s}(x),
\end{equation}
where $\mathbf{1}$ denotes the indicator function. As a direct implication of Campbell's theorem \citep{ChiuEtAl2013} the mean number of distance calculations per point is therefore
\begin{equation}\label{eqn:mean:n1}
\overline{n}_1=\mathsf{E}[n_1(x)]=\lambda\mathsf{E}[V_d(E_t(M,w))].    
\end{equation}

During the second step of the algorithm for each point $x\in X$ outside the union of all $s+E_t$, $(s,E_t)\in\Psi_t$, distances have to be calculated with respect to all marked points from $\Psi$ which are considered relevant. Practically this means that we fix some $R\geq0$ and restrict $\Psi$ to those generators $(s,M,w)$ whose seed point $s$ is contained in $W\oplus RB^d$, where $W\supset X$ is convex and compact. Denote this restriction by $\Psi^{(R)}$. If $R=0$, this means that only seeds within $W$ are considered; in the case $R=\sqrt{t_{\max}+w_{\max}}\,r_{\max}$ for a $0<t_{\max}<\infty$, (at least) all $(s,M,w)$ are considered for which $s+E_{t_{\max}}(M,w)\cap W\neq\varnothing$ applies. $\Psi^{(R)}$ is again a marked Poisson process.

Furthermore, assume $t\leq t_{\max}$ and let
\begin{equation}
\Psi^{(x,t)}=\{(s,M,w)\in\Psi:\,x\in s+E_t(M,w)\}.    
\end{equation}
$\Psi^{(x,t)}$ and $\Psi^{(R)}\setminus\Psi^{(x,t)}$ are then independent marked Poisson processes \citep{SchneiderWeil2008,ChiuEtAl2013}. Based on this we can estimate the mean number of distance calculation per point during the second step of the rendering algorithm.

For each point $x\in X$ the number of distance calculations during the second step is equal to
\begin{align}
\begin{split}
n_2(x)&=\mathbf{1}\{|\Psi^{(x,t)}|=0\}\cdot|\Psi^{(R)}|=\mathbf{1}\{|\Psi^{(x,t)}|=0\}\cdot(|\Psi^{(R)}\setminus\Psi^{(x,t)}|+|\Psi^{(x,t)}|)\\
&=\mathbf{1}\{|\Psi^{(x,t)}|=0\}\cdot|\Psi^{(R)}\setminus\Psi^{(x,t)}|,    
\end{split}
\end{align}
where $|\cdot|$ denotes cardinality. Due to the independence and the Poisson property of $\Psi^{(x,t)}$ and $\Psi^{(R)}\setminus\Psi^{(x,t)}$ we end up with
\begin{align}
\begin{split}
\overline{n}_2&=\mathsf{E}[n_2(x)]=\mathsf{E}\big[\mathbf{1}\{|\Psi^{(x,t)}|=0\}\cdot|\Psi^{(R)}\setminus\Psi^{(x,t)}|\big]=\mathsf{E}[\mathbf{1}\{|\Psi^{(x,t)}|=0\}]\cdot\mathsf{E}[|\Psi^{(R)}\setminus\Psi^{(x,t)}|]\\
&=\mathrm{e}^{-\mathsf{E}[|\Psi^{(x,t)}|]}\cdot\big(\mathsf{E}[|\Psi^{(R)}|]-\mathsf{E}[|\Psi^{(x,t)}|]\big)\\
&=\mathrm{e}^{-\lambda\mathsf{E}[V_d(E_t(M,w))]}\cdot\big(\overline{n}_R-\lambda\mathsf{E}[V_d(E_t(M,w))]\big)=\mathrm{e}^{-\overline{n}_1}\big(\overline{n}_R-\overline{n}_1\big),
\end{split}
\end{align}
where
\begin{equation}
\overline{n}_R=\mathsf{E}[|\Psi^{(R)}|]=\lambda V_d(W\oplus RB^d).
\end{equation}

In total, these are on average
\begin{equation}\label{eqn:n}
\overline{n}=\overline{n}_1+\overline{n}_2=\overline{n}_1+\mathrm{e}^{-\overline{n}_1}\big(\overline{n}_R-\overline{n}_1\big)
\end{equation}
distance calculations per point. Since $\overline{n}_R$ does not depend on $t$, $\overline{n}$ is minimized if it is minimized with respect to $\overline{n}_1$. A calculation shows that the optimum $\overline{n}_{1,\min}$ is given by
\begin{equation}\label{eqn:minimizer}
\mathrm{e}^{\overline{n}_1}+\overline{n}_1=\overline{n}_R+1. 
\end{equation}
Since $\overline{n}_1$ depends monotonically on $t$, an optimal $t$ can be determined using (\ref{eqn:minimizer}) if necessary.

From (\ref{eqn:minimizer}) we can further conclude that the optimal $\overline{n}_{1,\min}$ satisfies
\begin{equation}\label{eqn:upper}
\overline{n}_{1,\min}\leq\log(\overline{n}_R+1).
\end{equation}
Finally, substituting (\ref{eqn:minimizer}) into (\ref{eqn:n}) yields the following estimate for the minimum value of $\overline{n}$:
\begin{equation}
\overline{n}_{\min} = \overline{n}_{1,\min}+1-\mathrm{e}^{-\overline{n}_{1,\min}}\leq\log(\overline{n}_R+1)+1.
\end{equation}
The total complexity then equals $|X|(\log(\overline{n}_R+1)+1)$ and is thus of the order $\mathcal{O}(|X|\log(\overline{n}_R))$. The total complexity of the 'brute force' approach would be $\mathcal{O}(|X|\overline{n}_R)$.

The useful result of these considerations is that a value for $t$ can be determined from (\ref{eqn:minimizer}) that ensures this improved complexity and, at least on average, leads to a lower number of distance calculations and better overall rendering performance.

%
\subsection{Completed complexity estimation}
If we count the distance calculations within $B_t$ instead of $E_t$, the complexity class remains unchanged if we assume that there is a constant $c$, $1\leq c<\infty$, with which the ratio of $\mathsf{E}[V_d(B_t)]$ to $\mathsf{E}[V_d(E_t)]$ can be estimated from above. Indeed, under the above assumptions $w\leq w_{\max}<\infty$ and $\{x\in\mathbb{R}^d:\,x^{\top} Mx\leq1\}\subseteq r_{\max}B^d$ for $0<r_{\max}<\infty$, we have $B_t(M,w)\subseteq\sqrt{t+w_{\max}}\,\{y\in\mathbb{R}^d:\, |y_i|\leq r_{\max},\,i=1,\ldots,d\}$, which implies $\mathsf{E}[V_d(B_t)]\leq v_{\max}<\infty$, $v_{\max}=(2r_{\max}\sqrt{t+w_{\max}})^d$. Since $M$ is assumed to be positive definite and $t>0$, it also holds that $V_d(B_t)\geq V_d(E_t)>0$ and therefore $\mathsf{E}[V_d(B_t)]\geq\mathsf{E}[V_d(E_t)]=v_{\min}>0$. Together, this ensures that $c:=v_{\max}/v_{\min}$ is a possible such constant.

Then the mean number of distance calculations per point during the first step is not greater than $c\cdot\overline{n}_1$. Since nothing changes for the second step compared to \ref{subsec:simplified}, the mean total number $\overline{n'}$ of distance calculations per point can be estimated by
\begin{equation}\label{eqn:n:completed}
\overline{n'}\leq c\cdot\overline{n}_1+\mathrm{e}^{-\overline{n}_1}\big(\overline{n}_R-\overline{n}_1\big).
\end{equation}
The right-hand side of the inequality (\ref{eqn:n:completed}) is minimal for
\begin{equation}
c\,\mathrm{e}^{\overline{n}_1}+\overline{n}_1=\overline{n}_R+1.
\end{equation}
From this, we can conclude that the optimal $\overline{n}_1$ (for the right-hand side of (\ref{eqn:n:completed})) is not greater than $\log(\overline{n}_R+1) -\log(c)$ and 
that the minimum mean number of distance calculations per point is not greater than $c(\log(\overline{n}_R+1)+1-\log(c))$. Therefore, the total complexity is again of the order of $\mathcal{O}(|X|\log(\overline{n}_R))$.

%
\section{Conclusion}\label{sec:conclusion}
The aim of this article was to compile some further properties of GBPDs with regard to affine transformations and flat sections, which have not yet been explicitly discussed in the literature and are likely to play a role in future stochastic modelling with GBPDs. A simple conclusion from \ref{subsec:distortion} together with the mapping theorem for Poisson point processes is, for example, that for a marked stationary Poisson process $\Psi$ of intensity $\lambda$, as in Section \ref{sec:Poisson}, the transformed GBPD $A\mathcal{G}(\psi)$ for some invertible $d\times d$ matrix $A$ is also generated by a marked stationary Poisson process $A\Psi$ whose intensity has the value $\lambda/|\det(A)|$. Furthermore, it was demonstrated how an algorithm that is improved compared to a brute force approach can be applied to GBPDs, and the improved complexity was proven for the example of a marked stationary Poisson process. 

\bibliographystyle{apalike-ejor}
\bibliography{gbpd.bib}

\end{document}